# Low loss metasurfaces based on quantized meta-atom

Yisheng Gao

**Abstract**

Metasurfaces have been proposed as a new paradigm to manipulate light and improve light-matter interactions. Conventional metasurfaces are restricted to the loss of materials, limiting their performance ceiling. Here, the loss of metallic metasurfaces is reduced to the level of dielectric metasurfaces by quantizing the meta-atom, and a coupling model is proposed to interpret the loss reduction. The theoretical results are in excellent agreement with the finite-element method simulation results. This phenomenon is also reproduced in dielectric and second-quantized meta-atom metasurfaces. There is reason to believe that this is something new in the resonance field, and this work will shed new light on the designs and applications of metasurfaces. It will also pave theory support to other resonance fields beyond metasurfaces.

**Introduction**

Metasurfaces have emerged as one of the most important discoveries of this century, which allow us manipulate light in a way that is unattainable with naturally available materials. In the past few decades, some novel physical phenomena, including negative refractive index [1] and anomalous reflection and refraction [2], have been discovered in metasurfaces. Besides, various classical physical phenomena have been revealed in this system, such as Fano resonance [3, 4], electromagnetically induced transparency (EIT) [5, 6], bound states in the continuum (BIC) [7, 8], photonic Dirac system [9, 10] and so on. Those phenomena have been used to realize various applications, such as enhancing nonlinear effect [11, 12], special beams [13, 14], nanoprinting [15, 16], hologram [17, 18], metalens [19-21], optical switch [22, 23], quantum information processing [24, 25], optical data storage [26], etc. Those applications were designed with different materials and structures, two eternal subjects in metasurfaces [27]. Though various metals and dielectrics offered by nature inspire humankind to engineer and produce various metasurfaces with multifunction, the inconvenient truth is that the intrinsic loss of materials is ubiquitous and unavoidable, inhibiting the performance of metasurfaces, especially metallic ones [28]. To overcome the metal intrinsic loss, various complex dielectric and metal hybrid materials have been used [29, 30], but the setup is extremely complicated and introduces instability and noise [31, 32]. Nowadays, based on currently available information, though various structures have been employed in metasurfaces, there is still no efficient strategy to overcome the material intrinsic losses. Besides, though the optical properties of metasurfaces can be precisely simulated by the finite-element method (FEM) or finite-difference time-domain (FDTD) analysis, there is still a lack of a physical model or equations to describe the metasurface system. At present, the progress in this field relies on continuing breakthroughs in new materials, skilled fabrications, and cooperation with other conceptions. Something new is desired to enrich the metasurface system.

In analogy to the collision of ions, which have brought us much new physic and revealed the nature of existence, the meta-atom is opened to tackle these challenges. By quantizing the meta-atom, the loss of metallic metasurfaces is reduced to the level of dielectric metasurfaces. Based on the classical resonance in the harmonic oscillator system, a coupling model is proposed to describe normal metasurfaces and extended to the quantized meta-atom (QMA) metasurfaces. By solving the equations deduced from the model, the theoretical results are in excellent agreement with the

Email: yishngao@outlook.com

FEM simulation results. As a further verification, this phenomenon is reproduced in dielectric QMA metasurfaces and second QMA metasurfaces (the QMA are composed of another QMA).

**Schematic of QMA metasurfaces**

The basic concept underpinning the resonant metasurface is created by a 60-nm square lattice of Au squares. The whole structure is placed on a glass substrate ($n_s$ = 1.52, where $n$ is the refractive index) and coated with the same material ($n_c$ = 1.52). The width of the Au squares is $w$ = 519 nm, and the lattice spacing is $p$ = 1315 nm. The QMA is produced by splitting the meta-atom (Au square) into $m$ pieces in the x and y directions (Fig. 1A, take the case of $m$ = 3), which means each QMA is composed of $m \times m$ sub-meta-atoms (SMAs). Here, $m$ is defined as the quantum number. Inside the QMA, the ratio of the gap to the width of the SMA is $s$ ($s = (w_3 - w_3'): w_3'$, up inset in Fig. 1A, take the case of $s$ = 0.2). Figure 1B and C show the simulated transmittance spectra of the normal metasurface and QMA metasurface excited by an x-polarized field. After the meta-atom quantization, the QMA metasurface shows a sharper resonance along with a higher and flatter background than the normal metasurface. The full width at half maximum (FWHM) of the QMA metasurface transmission spectrum exhibits a decrease of 93.58% from 351.67 to 22.57 nm, reaching the level of the Si metasurface under the same conditions [33]. Meanwhile, the resonance position blue-shifts from 2277 to 2030 nm, and the minimum transmittance increases from 0.007 to 0.025. Besides, the electric field distribution at the resonance position changes from around to inside the meta-atom and concentrates in the gap of SMAs (insets in Fig. 1B and C), which provides a platform for enhancing the interaction between light and materials with a two orders enhancement [33].

**Optical properties of QMA metasurfaces**

Figure 1A demonstrates that the quantum number ($m$) and the gap ratio ($s$) serve as the primary parameters of QMA metasurfaces. Figure 2 presents the simulated transmittance spectra and the electric field pattern at the resonance position of the QMA metasurfaces with carious $m$ and $s$, when excited by an x-polarized field. Figure 2A displays the simulated transmittance spectra of the QMA metasurfaces with various $m$ but a fixed $s$. Figure 2B shows the minimum transmittance, resonance position, and FWHM of those spectra. As the quantum number varies from 2 to 9, the resonance position blue-shifts from 2043 to 2018 nm, and the FWHM decreases from 39.45 to 11.23 nm, while the minimum transmittance preserves below 0.06, respectively. Furthermore, as the quantum number increases, the three values exhibit a slower and slower rate of variation. Meanwhile, with the increase of the quantum number, the electric field at the resonance position becomes more concentrated inside the meta-atom, increasingly resembling a dielectric metasurface (Fig. 2C). Figure 2D displays the simulated transmittance spectra of the QMA metasurface with different $s$ but a fixed $m$. Figure 2E shows the minimum transmittance, resonance position, and FWHM of those spectra. As the gap ratio varies from 0.2 to 1.0, the resonance position blue-shifts from 2030 to 2011 nm, and the FWHM changes from 22.88 to 11.45 nm, while the minimum transmittance increases linearly from 0.025 to 0.16, respectively. Furthermore, the resonance position and FWHM also exhibit a slower and slower variation with the increase of the gap ratio. Figure 2F shows the electric field distribution at the resonance position. A boundary between adjacent QMAs occurs and becomes more and more apparent with the increased gap ratio.

The simulation results show that although the normal meta-atom is split into several SMAs, they can still work as a whole meta-atom. This is because the powerful electromagnetic field among SMAs can provide effective energy transfer and make the discrete SMAs whole [33]. Meanwhile, the lossless material in the gap can, in some sense, decrease the loss of the meta-atom, and the loss reduction also increases as the gap increase (Fig. 2D). However, the loss reduction is far from the gap ratio value. It implies that there are other mechanisms ruling this phenomenon.

**Mechanism of QMA metasurfaces**

To explore the mechanism of QMA metasurfaces, a model of two coupled driven oscillators [4] is adopted to describe normal metasurfaces first. The resonance of normal metasurfaces is a result of the coupling between the meta-atom resonance and the lattice resonance [34]. The relation matrix of metasurfaces can be described as the down panel of Fig. 3A. And the coupling of metasurfaces can be described by the matrix equation 1.

$$\begin{pmatrix} \varphi_1 & g_0 \\ g_0 & \varphi_0 \end{pmatrix} \begin{pmatrix} x_1 \\ x_0 \end{pmatrix} = i \begin{pmatrix} f_1 \\ f_0 \end{pmatrix} \tag{1}$$

$$\varphi_1 = \omega_1 - \omega - i\gamma_1, \quad \varphi_0 = \omega_0 - \omega - i\gamma_0$$

Here, $x_{i,i=1,2}$, $\omega_{i,i=1,2}$, and $\gamma_{i,i=1,2}$ are the oscillator amplitudes, resonant frequencies, and damping coefficients, respectively. $f_{i,i=1,2}$ are the external forces with the driving frequency $\omega$, $g_0$ is the coupling coefficient between the meta-atom resonance and the lattice resonance. The amplitudes of meta-atom oscillators ($x_1$) are obtained by solving the equation 1 [33]. And then, the theoretical meta-atom amplitudes ($|x_1|^2$) are compared with the simulated electric field intensity of meta-atom ($|Ex|^2$) [33]. The comparison results show that the built model can describe the metasurfaces system well [33].

Based on the model of normal metasurfaces, the QMA oscillator replaces the meta-atom oscillator to explain the mechanism of QMA metasurfaces (Fig. 3B, take the case of $m = 3$). When only considering the coupling of the adjacent SMAs, the relation matrix of QMA metasurface can be described as the down panel of Fig. 3B. Where, $\varphi_3 = \omega_3 - \omega - i\gamma_3$, $\omega_3$ and $\gamma_3$ are the resonant frequencies and damping coefficient of the SMA resonance, respectively. $\eta_3$ is the coupling coefficient of the adjacent SMAs. $g_0$ is the coupling coefficient between the QMA resonance and the lattice resonance. Actually, every SMA has an individual coupling coefficient ($g_{i,i=1,2,3}$) to the lattice oscillator amplitude ($x_0$), as shown in Fig. 3C. Unfortunately, it is hard to evaluate the coupling coefficient for every SMA. To overcome this trouble, different lattice resonance amplitude with the same coupling coefficient ($x_{i,i=4,5,6}, g_0$, Fig.3D) is used to replace the same lattice oscillator amplitude with different coupling coefficient ($x_0, g_{i,i=1,2,3}$, Fig.3C). The QMAs metasurfaces ($m = 3$) system can be described as the matrix equation 2:

$$\begin{pmatrix} \varphi_3 & \eta_3 & 0 & g_0 & 0 & 0 \\ \eta_3 & \varphi_3 & \eta_3 & 0 & g_0 & 0 \\ 0 & \eta_3 & \varphi_3 & 0 & 0 & g_0 \\ g_0 & 0 & 0 & \varphi_0 & 0 & 0 \\ 0 & g_0 & 0 & 0 & \varphi_0 & 0 \\ 0 & 0 & g_0 & 0 & 0 & \varphi_0 \end{pmatrix} \begin{pmatrix} x_1 \\ x_2 \\ x_3 \\ x_4 \\ x_5 \\ x_6 \end{pmatrix} = i \begin{pmatrix} f_1 \\ f_2 \\ f_3 \\ f_4 \\ f_5 \\ f_6 \end{pmatrix} \tag{2}$$

The theoretical equations for other QMA metasurfaces with different quantum numbers can be derived using the same method [33]. Meanwhile, As the connecting electromagnetic field of the adjacent SMAs is radiation-induced by SMAs, the coupling coefficient $\eta_m$ should be related to the damping coefficient $\gamma_m$; Here, define $\eta_m = v\gamma_m$, $v$ is the relative damping coefficient. For simplicity, assume the resonance frequency and the damping coefficient of SMAs: $\omega_m = m\omega_1$, and $\gamma_m = m\gamma_1$, respectively.

**The simulation and theoretical results of QMA metasurfaces**

To verify the built model, Figure 4 compares the simulation and theoretical results with different quantum numbers and gaps (*s* for Sim. *v* for Theo.). The theoretical calculations are performed with the same parameters of equation 1, except a fixed blueshift 193 nm for the lattice resonance [33]. Subsequently, the squared modulus of the simulated volume average electric field ($|Ex|^2$) in the original meta-atom region is compared with the squared modulus of the theoretical amplitude of SMAs ($|x|^2, x = \sum_{i=1}^{m} x_i$) [33].

Figure 4A shows the simulated electric field intensity spectra of SMAs with *m* = 2, 3, 5, 7, 9 and *s* = 0.2. Figure 4B shows the theoretical intensity spectra of SMAs with *m* = 2, 3, 5, 7, 9 and *v* = -0.7. With the increase of the quantum number, the theoretical spectra have a similar variation trend as the simulated spectra. Figure 4C, D, and E show the maximum intensity, resonance position, and FWHM of the theoretical (red curve) and simulated (black curve) spectra as a function of the quantum number, respectively. The theoretical results exhibit a larger intensity than the simulation as the quantum number increases. The difference mainly comes from the fixed coupling coefficient ($g_0$) and lattice resonance loss ($\gamma_0$), which are used in the theoretical calculation [33]. Additionally, the theoretical results exhibit a minor blueshift of the resonance position than the simulation. It is because the theoretical calculations are carried out with a fixed blueshift of the lattice resonance, yet the blueshift of the lattice resonance increases with the increase of the quantum number [33]. Meanwhile, the simulation and theoretical result demonstrate almost identical FWHM as the gap increases, further validating the initial assumption of the SMA ($\omega_m = m\omega_1, \gamma_m = m\gamma_1$).

Figure 4F shows the simulated electric field intensity spectra of SMAs with *m* = 3 and *s* = 0.2, 0.3, 0.4, 0.6, 1.0. Figure 4G shows the theoretical intensity spectra of SMAs with *m* = 3 and *v* = -0.7, -0.4, -0.1, 0.3, 0.9. Figure 4H, I, and J show the maximum intensity, resonance position, and FWHM of the theoretical (red curve) and simulated (black curve) spectra as a function of the gap, respectively. The red solid circle and black solid square in Fig. 4H, I, and J are the data from Fig. 4F and G, respectively. The real gap (x-axis value) is calculated by $w(1 + s)/3s$. The simulation results show that the maximum electric field intensity increases first, then decreases as the gap increases (black curve in Fig. 4H). The theoretical results also show this trend when the relative damping coefficient increased from negative to positive (red curve in Fig. 4H). Surprisingly, the theoretical results demonstrate incredible conformance with the simulation before the highest point (*s* = 0.4, *v* = -0.1). Nevertheless, the theoretical results show a slightly bigger reduction rate after the highest point than the simulation. Those results imply that gain and loss mechanisms rule the coupling of the adjacent SMAs, like the repulsion and gravitation in atoms [33]. Besides, the simulated resonance position shows a nonlinear blueshift with the increase of the gap, yet the theoretical results show an almost linear blueshift. It implies that the increase of gap not only introduces the variation of damping coupling coefficient but also affects other elements ($g_0, \Delta\lambda_0$), which can cause the nonlinear blueshift of the resonance position [33]. Additionally, the simulation and theoretical results have almost identical FWHM as the gap increases, similar to the different quantum numbers case.

**Dielectric QMA metasurfaces and second QMA metasurfaces**

Figure 4 demonstrates that the built model can describe metallic QMA metasurfaces well. Moreover, as the model is based on the classical resonance system, the loss reduction phenomenon from the quantizing meta-atom should work in other metasurfaces. As a further verification, a dielectric QMA metasurface and second QMA metasurface are built and simulated, as shown in Fig. 5.

The basic dielectric metasurface is composed of 110-nm Si squares with a width of 796 nm and lattice spacing of 1315 nm, and the dielectric QMA metasurface is built by quantizing the meta-atom with the parameters $m = 3$ and $s = 0.2$. Figure 5A shows the simulated transmittance spectrum of the Si normal (black curve) and QMA (red curve) metasurfaces excited by an x-polarized field. The top and down insets show the electric field pattern of the normal and QMA metasurfaces at the resonance position, respectively. After the quantization, the Si QMA metasurface shows a similar variation trend with the Au QMA metasurfaces case (Fig. 1B and C). The FWHM of the Si QMA metasurface transmission spectrum exhibits a decrease of 78.91% from 47.47 to 10.01 nm, the resonance position blue-shifts from 2084 to 2020 nm, and the minimum transmittance increases from 0.002 to 0.019. Besides, the electric field distribution also becomes concentrated in the gap of SMAs (insets in Fig. 5A).

The second QMA metasurface is created by quantizing the SMAs of a QMA metasurface with the parameters $m'$ and $s'$ ($s' = (w_{3\_3} - w'_{3\_3}): w'_{3\_3}$, Fig. 5B). Figure 5C shows the simulated transmittance spectrum of the QMA metasurface (red curve) and its corresponding second QMA metasurface ($m' = 3$, $s' = 0.1$, black curve), excited by an x-polarized field. The QMA metasurface is the QMA metasurface in Fig. 1C. The top and down insets show the electric field pattern of the QMA and second QMA metasurfaces at the resonance position, respectively. The quantization of the SMAs also introduces a similar variation trend with the Au QMA metasurface case (Fig. 1B and C). The FWHM of the second QMA metasurface transmission spectrum exhibits a decrease of 64.69% from 22.57 to 7.97 nm, and the resonance position blue-shifts from 2030 to 2009 nm. Nonetheless, the minimum transmittance increases from 0.025 to 0.237, resulting from the weaken electromagnetic among the quantized SMAs (insets in Fig. 5C). Additionally, the electric field distribution becomes more concentrated in the internal gap of the quantized SMAs (insets in Fig. 5C) [33].

**Conclusion**

This work has solved the long-held intrinsic loss problem in metallic metasurface that is almost unsolvable by existing know-how. Besides, this work verifies that a coupling model can illustrate the physics of metasurfaces well. Moreover, the proposed meta-atom quantization will empower unprecedented advances and broaden the applications in metasurfaces. Furthermore, this work can also be extended to other resonance systems, since the theoretical model is based on classical resonances, which signifies the loss reduction from quantizing the meta-atom should be something new in the resonance system beyond metasurfaces.

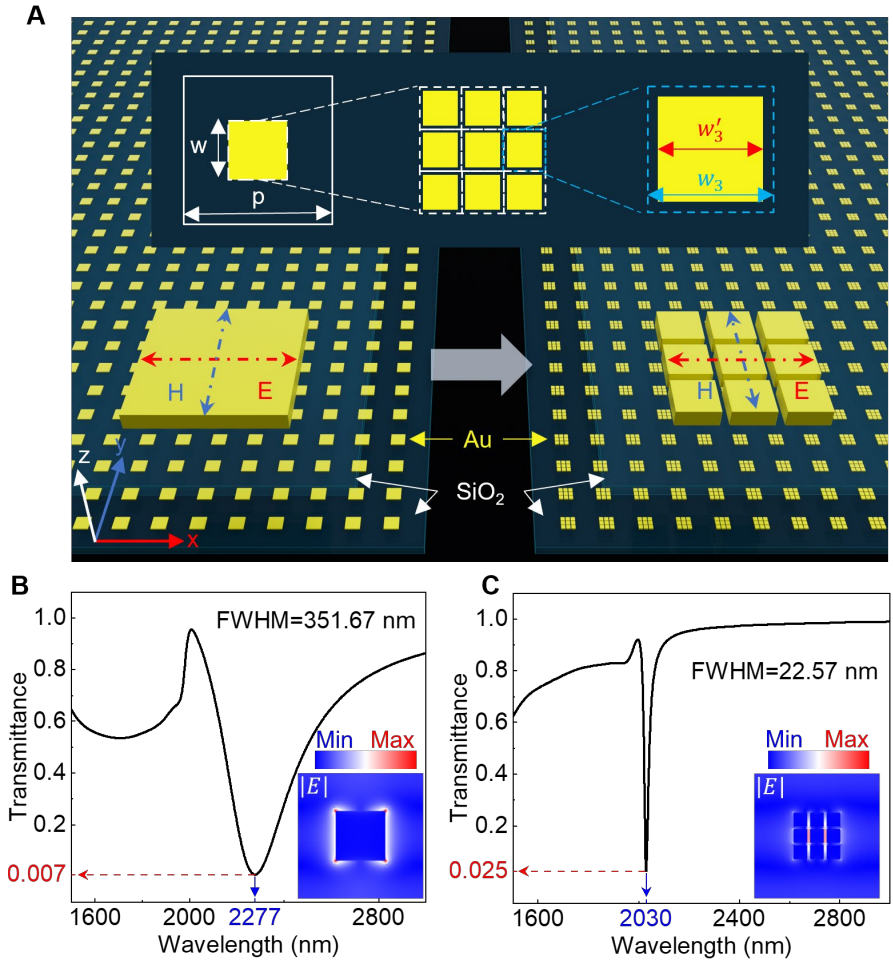

**Fig. 1. Schematic of QMA metasurfaces.** (**A**) Schematic of the normal and QMA metasurface. The normal metasurface is created by a square lattice of Au squares (meta-atom) surrounded by the same material (silica). The QMA is produced by splitting the meta-atom into $m$ pieces in the x and y directions with a gap ratio of $s$. The top panel inset (dark blue area) shows the process of quantizing the meta-atom, taking the case of $m = 3$, $s = (w_3 - w'_3):w'_3 = 0.2$. (**B**) Simulated transmittance spectrum of the normal metasurface excited by an x-polarized field. The inset shows the electric field pattern of the normal metasurface at the resonance position. The thickness, width, and lattice spacing of the Au square are, respectively, $h = 60$ nm, $w = 519$ nm, and $p = 1315$ nm. (**C**) Simulated transmittance spectrum of the QMA metasurface excited by an x-polarized field. The inset shows the electric field pattern of the QMA metasurface at the resonance position. The QMA is produced by quantizing the Au square in (B) with the parameters $m = 3$ and $s = 0.2$.

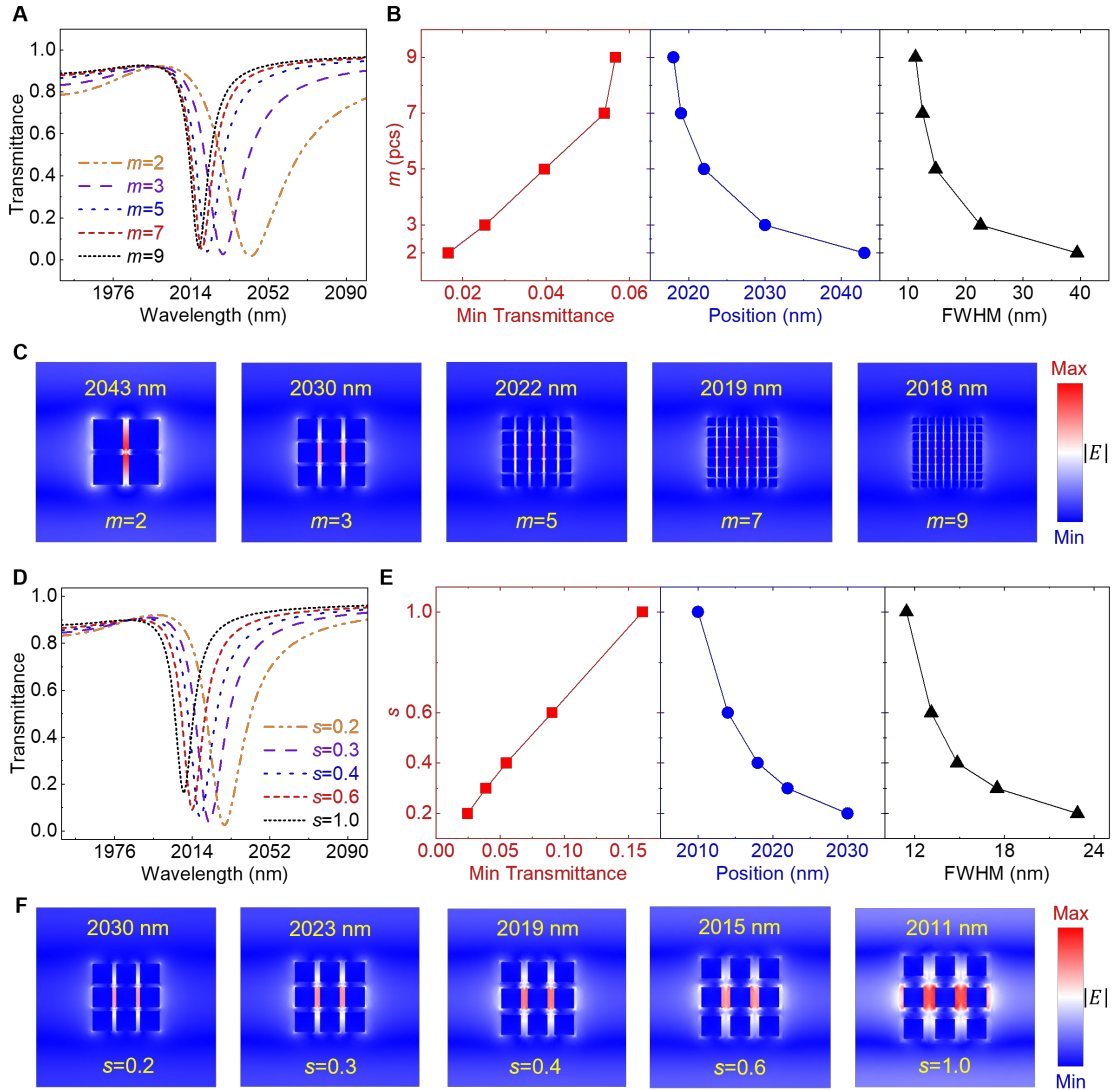

**Fig. 2. Characterization of QMA metasurfaces. (A)** Simulated transmittance spectra of the QMA metasurfaces with $m$ = 2, 3, 5, 7, 9 and $s$ = 0.2, excited by an x-polarized field. **(B)** The minimum transmittance, resonance position, and FWHM of the transmittance spectra in (A). **(C)** Electric field pattern at the resonance position of the QMA metasurfaces in (A). **(D)** Simulated transmittance spectra of the QMA metasurfaces with $m$ = 3 and $s$ = 0.2, 0.3, 0.4, 0.6, 1.0, excited by an x-polarized field. **(E)** The minimum transmittance, resonance position, and FWHM of the transmittance spectra in (C). **(F)** Electric field pattern at the resonance position of the QMA metasurfaces in (C).

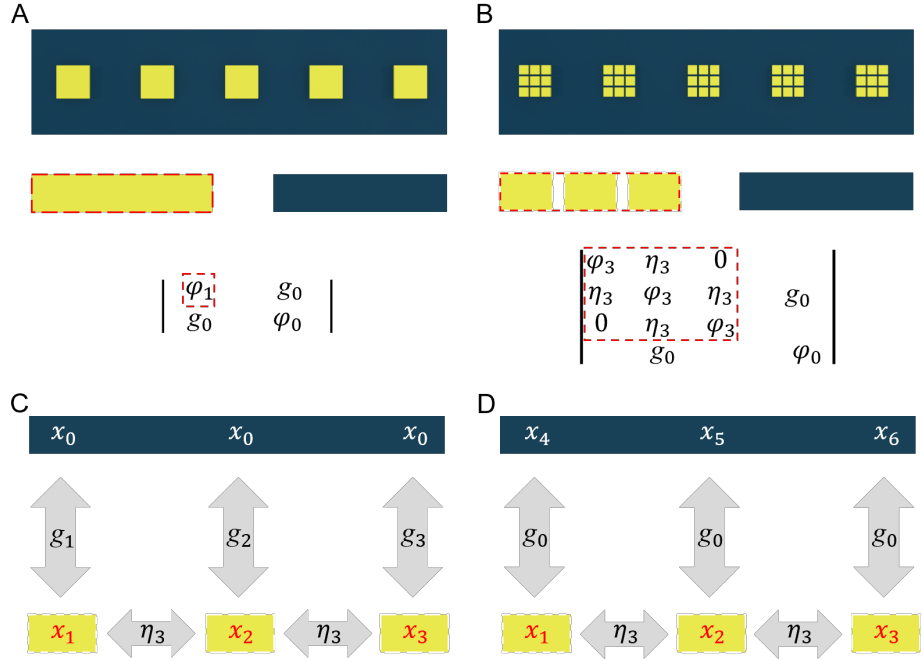

**Fig. 3. Mechanism of QMA metasurfaces.** (**A** and **B**) Respectively, the diagram (up panel) and relation matrix (down panel) of normal and QMA metasurfaces. Here, $g_0$ is the coupling coefficient between the meta-atom and the lattice resonance. $\eta_3$ is the coupling coefficient between the adjacent SMAs. $\varphi_{m,m=0,1,3} = \omega_m - \omega - i\gamma_m$, $\omega_m$ is the resonant frequency, $\gamma_m$ is the damping coefficient, here, 0 for lattice resonance, 1 for normal meta-atom resonance, and 3 for SMA resonance. (**C** and **D**) Respectively, the actual diagram and equivalent diagram of the elements in QMA metasurfaces. Here, $x_{m,m=1,2,3}$ is the oscillator amplitude of each SMA. $x_0$ is the oscillator amplitude of lattice resonance. $g_{m,m=1,2,3}$ is the coupling coefficient between each SMA and lattice resonance. $x_{m,m=4,5,6}$ is the oscillator amplitude of equivalent lattice resonance corresponding to each SMA with the same coupling coefficient ($g_0$).

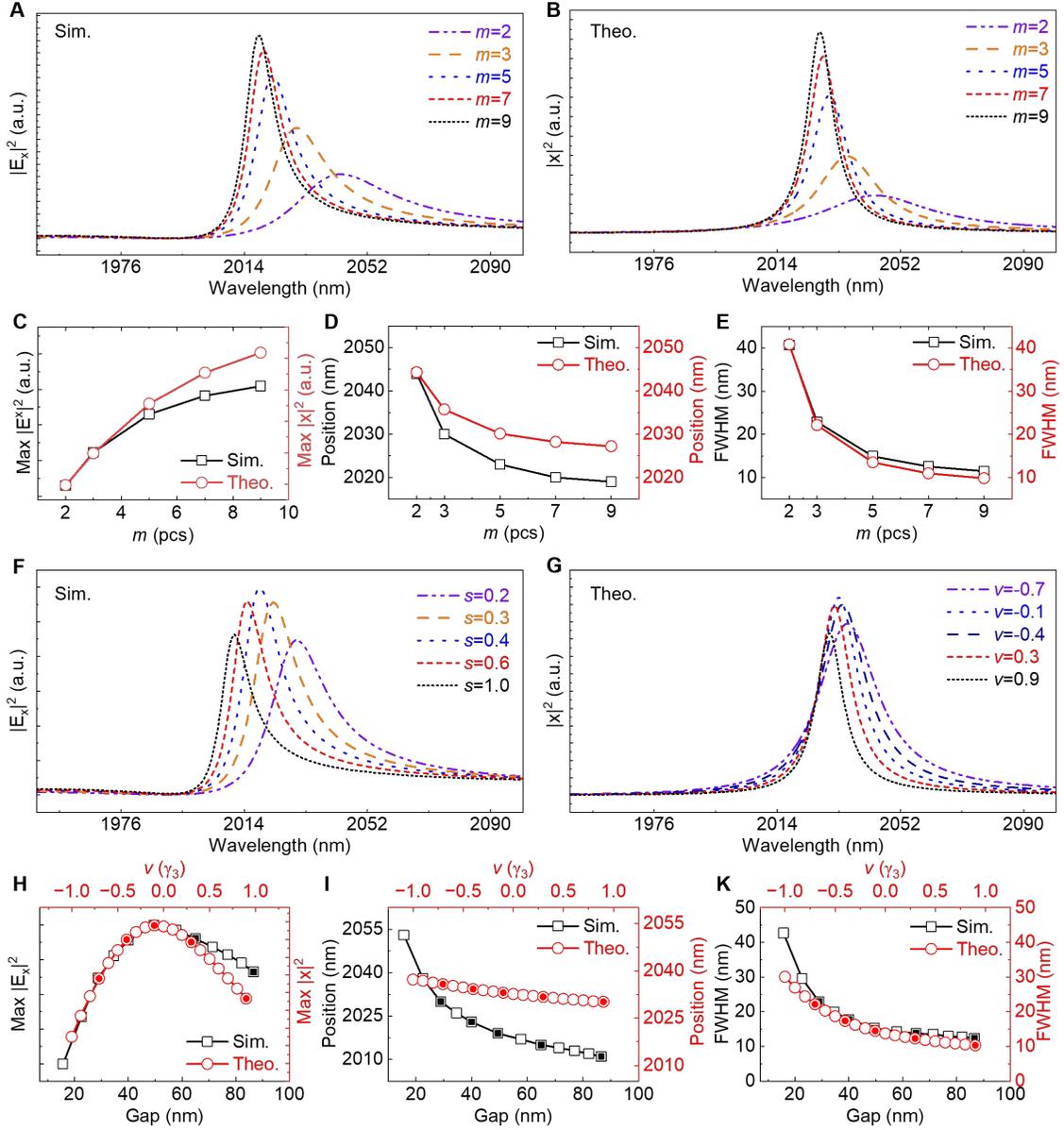

**Fig. 4. The simulation and theoretical results of QMA metasurfaces.** (**A**) The squared modulus of the simulated volume average electric field ($|Ex|^2$) in the original meta-atom region of the QMA metasurfaces with $m$ = 2, 3, 5, 7, 9 and $s$ = 0.2, excited by an x-polarized field. (**B**) The squared modulus of the theoretical amplitude ($|x|^2$, $x = \sum_{i=1}^{m} x_i$) of the QMA metasurfaces with $m$ = 2, 3, 5, 7, 9 and $v$ = -0.7. (**C**, **D**, and **E**) Respectively, the maximum intensity, resonance position, and FWHM of the simulated (black curve) and theoretical (red curve) intensity spectra of the QMA metasurfaces with $m$ = 2, 3, 5, 7, 9 and a fixed gap ($s$ = 0.2, $v$ = -0.7). (**F**) The squared modulus of the simulated volume average electric field ($|Ex|^2$) in the original meta-atom region of the QMA metasurfaces with $m$ = 3 and $s$ = 0.2, 0.4, 0.6, 1.0, excited by an x-polarized field. (**G**) The squared modulus of the theoretical amplitude ($|x|^2$, $x = \sum_{i=1}^{m} x_i$) of the QMA metasurfaces with $m$ = 3 and $v$ = -0.7, -0.4, -0.1, 0.3, 0.9. (**H**, **I**, and **J**) Respectively, the maximum intensity, resonance position, and FWHM of the simulated (black curve) and theoretical (red curve) intensity spectra of the QMA metasurfaces with $m$ = 3 and different gaps (Red solid circle: $s$ = 0.2, 0.4, 0.6, 1.0, black square: $v$ = -0.7, -0.4, -0.1, 0.3, 0.9). The x-axis value (real gap) is calculated by $w(1 + s)/3s$.

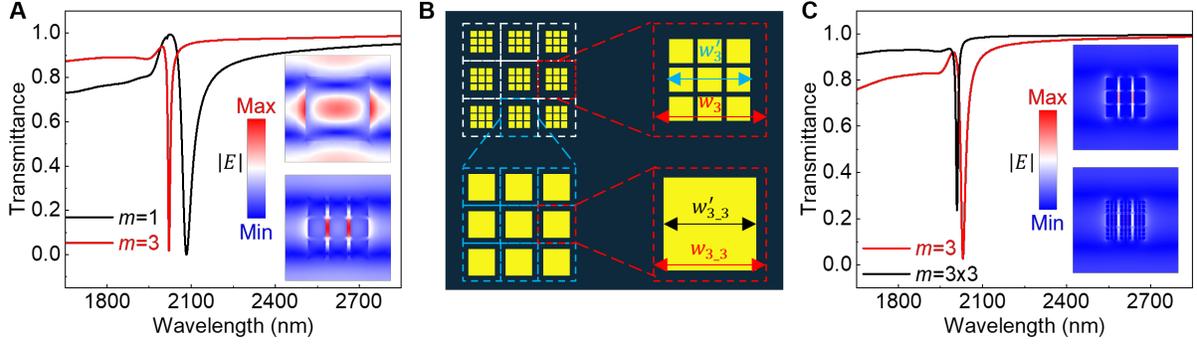

**Fig. 5. Dielectric QMA metasurfaces and second QMA metasurfaces.** (**A**) Simulated transmittance spectrum of the Si normal (black curve) and QMA (red curve) metasurfaces, excited by an x-polarized field. The top and down insets show the electric field pattern of the Si normal and QMA metasurfaces at the resonance position, respectively. The Si normal metasurface is composed of 110-nm Si squares with a width of 796 nm and lattice spacing of 1315 nm. The Si QMA metasurface is built by quantizing the Si squares with the parameters $m = 3$ and $s = 0.2$. (**B**) Schematic of second QMA metasurfaces. The second QMA metasurface is created by quantizing the SMA of a QMA metasurface with the parameters $m'$ and $s' = (w_{3\_3} - w'_{3\_3}) : w'_{3\_3}$. (**C**) Simulated transmittance spectrum of the QMA (red curve) and second QMA (black curve) metasurfaces, excited by an x-polarized field. The QMA metasurface is the QMA metasurface in Fig. 1C. The second QMA metasurface is created with the parameters $m' = 3$ and $s' = 0.1$, marked by $m = 3 \times 3$. The top and down insets show the electric field pattern of the QMA and second QMA metasurfaces at the resonance position, respectively.